\documentclass[aps,pra,11pt,longbibliography]{revtex4-1}
\usepackage[latin1]{inputenc} 
\usepackage[T1]{fontenc}
\usepackage{lmodern}
\usepackage{graphicx}
\usepackage{amsmath,amssymb,color,xcolor,bm}
\usepackage{soul}
\renewcommand{\d}{\mathrm{d}}

\newcommand{\q}{\mathbf{q}}

\renewcommand{\Im}[1]{\mathrm{Im} \, [ #1]}

\begin{document}
\title{Electron cooling in graphene enhanced by plasmon-hydron resonance}

\author{Xiaoqing Yu$^{1}$, Alessandro Principi$^2$, Klaas-Jan Tielrooij$^{3,4}$,  Mischa Bonn$^1$ and Nikita Kavokine$^{1,5}$}
\email{nikita.kavokine@mpip-mainz.mpg.de}
\affiliation{$^1$Max Planck Institute for Polymer Research, Ackermannweg 10, Mainz 55128, Germany}
\affiliation{$^2$School of Physics and Astronomy, University of Manchester, M13 9PL Manchester, U.K.}
\affiliation{$^3$Catalan Institute of Nanoscience and Nanotechnology (ICN2), BIST and CSIC, Campus UAB, Bellaterra, Barcelona, 08193, Spain}
\affiliation{$^4$Department of Applied Physics, TU Eindhoven, Den Dolech 2, 5612 AZ, Eindhoven, The Netherlands}
\affiliation{$^5$Center for Computational Quantum Physics, Flatiron Institute, 162 5$^{\rm th}$ Avenue, New York, NY 10010, USA}

\begin{abstract}
\textbf{
Evidence is accumulating for the crucial role of a solid's free electrons in the dynamics of solid-liquid interfaces. Liquids induce electronic polarization and drive electric currents as they flow; electronic excitations, in turn, participate in hydrodynamic friction. Yet, the underlying solid-liquid interactions have been lacking a direct experimental probe. Here, we study the energy transfer across liquid-graphene interfaces using ultrafast spectroscopy. The graphene electrons are heated up quasi-instantaneously by a visible excitation pulse, and the time evolution of the electronic temperature is then monitored with a terahertz pulse. We observe that water accelerates the cooling of the graphene electrons, whereas other polar liquids leave the cooling dynamics largely unaffected. A quantum theory of solid-liquid heat transfer accounts for the water-specific cooling enhancement through a resonance between the graphene surface plasmon mode and the so-called hydrons -- water charge fluctuations --, particularly the water libration modes, that allows for efficient energy transfer. Our results provide direct experimental evidence of a solid-liquid interaction mediated by collective modes and support the theoretically proposed mechanism for quantum friction. They further reveal a particularly large thermal boundary conductance for the water-graphene interface and suggest strategies for enhancing the thermal conductivity in graphene-based nanostructures. 
}
\end{abstract}

\maketitle

Free electrons in graphene exhibit rather unique dynamics in the terahertz (THz) frequency range, including a highly non-linear response to photoexcitation by THz pulses~\cite{Hwang2013,Hafez2018}. Graphene's distinctive dynamical properties on picosecond timescales have found several applications in, e.g., ultrafast photodetectors, modulators, and receivers~\cite{Liu2011,Romagnoli2018,Muench2019}. The THz frequency range acquires particular importance at room temperature $T$, where it corresponds to the typical frequency of thermal fluctuations: $k_{\rm B} T/ \hbar \sim 6 ~\rm THz$, with $k_{\rm B} $ Boltzmann's constant and $\hbar$ Planck's constant. One may therefore expect non-trivial couplings between the graphene electrons and the thermal fluctuations of their environment. These couplings have been intensively studied in the case of a solid environment: for instance, non-adiabatic effects have been shown to arise in the graphene electron-phonon interaction~\cite{Pisana2007}, and plasmon-phonon coupling between graphene and a polar substrate has been demonstrated~\cite{Hwang2010,Dai2015, Koch2016}. More recently, it has been theoretically proposed that similar effects are at play when graphene has a liquid environment: then, the interaction between the liquid's charge fluctuations -- dubbed hydrons -- and graphene's electronic excitations tunes the hydrodynamic friction at the carbon surface~\cite{Kavokine2022,Bui2022}. This "quantum friction" mechanism holds the potential of entirely new strategies for controlling liquid flows at nanometer scales~\cite{Coquinot2022,Marcotte2022}; {it is therefore of interest to experimentally probe the underlying electron-hydron interaction}. 

In this Article, {we probe solid-liquid interactions by measuring energy transfer at the solid-liquid interface}. Specifically, we use a femtosecond visible pulse to introduce a quasi-instantaneous temperature difference between the {electrons of a graphene sample} and their environment. The cooling rate of the electronic system is followed in real-time using terahertz pulses. Such Optical Pump - Terahertz Probe (OPTP) spectroscopy is a well-established tool for probing electron relaxation in 2D materials~\cite{George2008,Kar2015,Mihnev2015,Mihnev2016,Pogna2021,Zheng2022}. In high-quality graphene, it has been used to identify the interaction of hot carriers with optical phonons~\cite{Mihnev2016,Pogna2021} and with substrate phonons as the main electron cooling mechanisms~\cite{Tielrooij2018}; {it has also identified the role of Coulomb interactions in the interlayer thermal conductivity of graphene stacks~\cite{Mihnev2015}}. Here, we measure the electron relaxation time in the presence of different polar liquids to probe the electron-hydron interaction, which we find to be {comparable} to the electron - optical phonon interaction only when the liquid is water. A complete theoretical analysis shows that this specificity of water is explained by the strong coupling of its THz (libration) modes to the graphene surface plasmon, with the electron-electron interactions in graphene playing a crucial role. 

\begin{figure}
\centering
\includegraphics{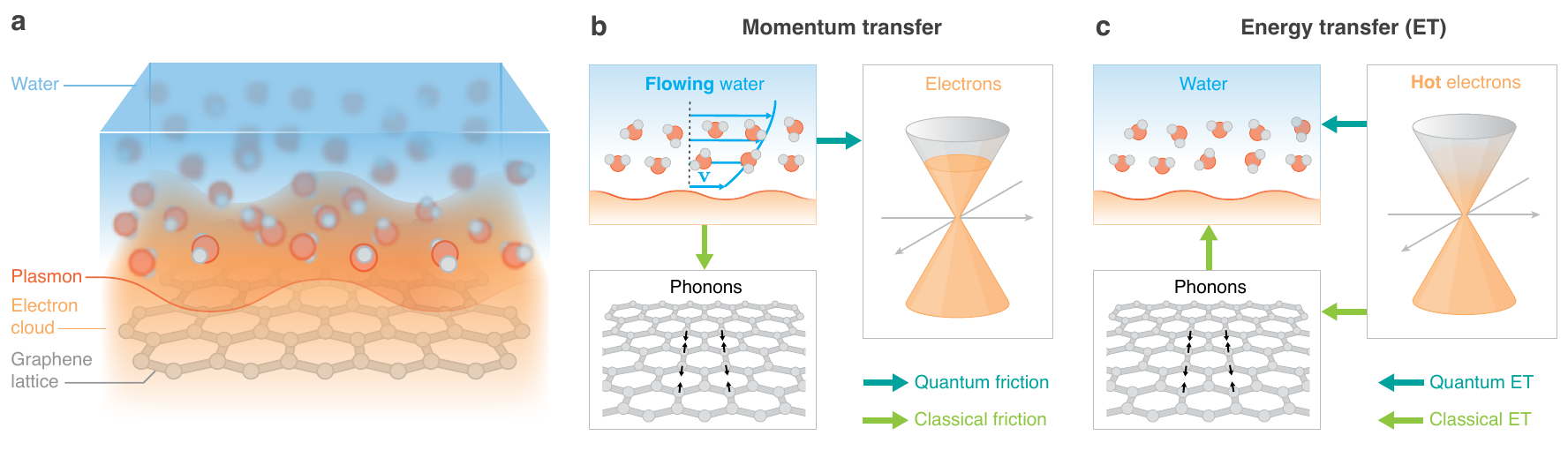}
\caption{\textbf{Heat transfer and friction at the solid-liquid interface}. \textbf{a}. Schematics of the system under study: the interface between water and a graphene sheet. {The picture emphasizes the electron cloud and its wave-like plasmon excitation.} 
{\textbf{b}. Momentum transfer processes at the solid-liquid interface. A flowing liquid (the flow profile is shown by the thin blue arrows) may not only transfer momentum to the crystal lattice (exciting phonon vibrations) through classical hydrodynamic friction, but also directly to the electrons through quantum friction. \textbf{c}. Energy transfer (ET) processes at the solid-liquid interface. In the typically-assumed "classical" pathway, hot electrons first transfer energy to the phonons, which transfer energy to the liquid. An alternative "quantum" pathway consists in the electrons transferring energy directly to the liquid through Coulomb coupling.}}
\end{figure}

\vskip0.5cm
\noindent{\bf \large Solid-liquid heat transfer} \\
The energy transfer between a solid and a liquid is usually considered to be mediated by molecular vibrations at the interface, as most of a solid's heat capacity is contained in its phonon modes~\cite{Phillpot2005}. Even if an optical excitation of the solid's electrons is used to create the temperature difference, the electrons are typically assumed to thermalize with phonons on a very short time scale, so that the solid's phonons ultimately mediate the energy transfer to the liquid's vibrational modes~\cite{Gutierrez2022, Herrero2022}. However, if the electrons were to transfer energy to the liquid faster than to the phonons, the interfacial thermal conductivity would contain a non-negligible contribution from near-field radiative heat transfer~\cite{Volokitin2007,Biehs2021} (Fig. 1c). Such an electronic or "quantum" contribution to heat transfer is in close analogy with the quantum contribution to hydrodynamic friction. Quantum hydrodynamic friction relies on momentum being transferred directly between the solid's and the liquid's charge fluctuation modes, {coupled by Coulomb forces (Fig. 1b): the two processes are mediated by the same solid-liquid interaction.} 

{In order to probe this interaction through a hydrodynamic friction measurement, one needs to ensure that quantum friction dominates over the classical surface roughness contribution: this imposes stringent constraints on the sample's surface state, in already technically-difficult experiments~\cite{Maali2008,Secchi2016,Xie2018}. Similarly, in the case of energy transfer, the quantum contribution needs to be comparable to the classical phonon-based contribution in order to become measurable; yet, this condition is easier to satisfy since it is insensitive to the sample's surface roughness.} We show that this condition is met upon optically exciting a graphene-water interface, owing, in particular, to graphene's weak electron-phonon coupling~\cite{Bistritzer2009,Betz2012}. 

 \begin{figure}
\centering
\includegraphics{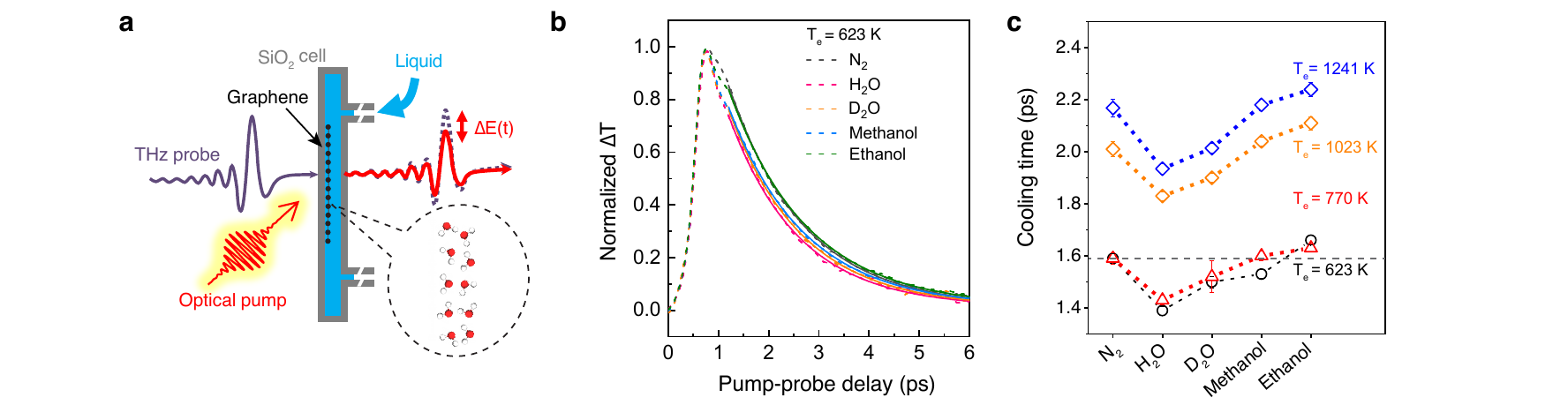}
\caption{\textbf{Measurement of picosecond hot electron relaxation in graphene}. \textbf{a}. Schematic of the experimental setup. A graphene sample {(Fermi level in the range $\rm 100 - 180~meV$, see SI Sec. 1.4)}  is placed in contact with a liquid inside a fused silica flow cell. An optical excitation pulse impulsively heats up the graphene electrons, and the electron temperature dynamics are then monitored with a THz probe. \textbf{b}. Normalized electron temperature as a function of time after photoexcitation. The dotted lines represent raw data and the full lines are exponential fits. \textbf{c}. Electron cooling time obtained through exponential fitting (see \textbf{b}) for the different liquids that have been placed in the flow cell and different initial electron temperatures, set by the excitation laser fluence. Faster cooling is observed in the presence of water and heavy water. Error bars represent 95\% confidence intervals of the exponential fits, and the centre point corresponds to the result of the least-squares fitting procedure.}
\end{figure}
 
\vskip0.5cm
\noindent{\bf \large Time-resolved electron cooling} \\
Our experimental setup is schematically represented in Fig. 2a. A monolayer graphene sample {grown by chemical vapor deposition (CVD)} was transferred onto a fused silica flow cell, filled with either nitrogen gas or a liquid of our choice (SI Sec. 1.1). {The graphene chemical potential was in the range $\rm 100-180~meV$ as determined from Raman measurements (SI Sec. 1.4).} In a typical experiment, the graphene electrons were excited using a $\sim 50 ~\rm fs$ laser pulse with 800 nm central wavelength. Then, the attenuation of a $\sim 1 ~\rm ps$ THz probe pulse (precisely, the modulation of the peak electric field) was monitored as a function of the pump-probe delay (SI Sec. 1.2). After absorption of the exciting pump pulse, the non-equilibrium electron distribution typically thermalizes over a sub-100 fs timescale through electron-electron scattering~\cite{Brida2013}: it can then be described as a Fermi-Dirac distribution at a given temperature. A hotter electron distribution results in a lower THz photoconductivity, since hotter electrons are less efficient at screening charged impurities~\cite{Tomadin2018,Massicotte2021}. The pump-probe measurement thus gives access to the electron temperature dynamics after photoexcitation (Fig. 2b). 

Regardless of the medium that the graphene is in contact with, the electronic temperature $T(t)$ exhibits a relaxation that can be approximated by an exponential function : $\Delta T(t) = T(t) - T_0 = \Delta T_0 e^{-t /\tau}$. This allows us to extract the cooling times $\tau$ for the different liquids and different initial electronic temperatures (determined by the excitation laser fluence), displayed in Fig. 2c. We observe that the cooling time is longer for an initially hotter electron distribution, in agreement with previous reports~\cite{Pogna2021}. Now, for all initial temperatures, we consistently observe the same dependence of the cooling time on the sample's liquid environment. In the presence of water ($\rm H_2O$) and heavy water ($\rm D_2 O$), the graphene electrons cool faster than they do intrinsically, in an inert nitrogen atmosphere. Conversely, methanol and ethanol have almost no effect on the electron cooling time. Interestingly, we observe an isotope effect in the electron cooling process: there is a difference in the cooling times in the presence of $\rm H_2O$ and $\rm D_2O$ that well exceeds experimental uncertainties. 

We are thus led to hypothesize, as anticipated above, that the liquid provides the electrons with a supplementary cooling pathway, which, in the case of water, has an efficiency comparable to the intrinsic cooling pathway via phonons. We then interpret the faster cooling as a signature of "quantum" electron-liquid energy transfer. We assess the pertinence of this hypothesis by developing a complete theory of quantum energy transfer at the solid-liquid interface.

\vskip0.5cm
\noindent{\bf \large Theoretical framework} \\
In order to tackle the interaction between a classical liquid and an electronic system whose behavior is intrinsically quantum, we describe the liquid in a formally quantum way. Following ref.~\cite{Kavokine2022}, we represent the liquid's charge density as a free fluctuating field with prescribed correlation functions. This naturally leads to a Fourier-space description of the solid-liquid interface in terms of its collective modes, rather than the usual molecular scale interactions. Within this description, the quantum solid-liquid energy transfer amounts to electron relaxation upon coupling to a bosonic bath, a problem that has been extensively studied in condensed matter systems~\cite{Mahan}. Interestingly, in the case of graphene, many of these studies are carried out within a single-particle Boltzmann formalism, which may incorporate multiple screening effects only in an \emph{ad hoc} fashion~\cite{Bistritzer2009,Principi2017,Pogna2021}. These effects turn out to be crucial for the solid-liquid system under consideration: we have therefore developed an \emph{ab initio} theory of solid-liquid heat transfer based on the non-equilibrium Keldysh formalism~\cite{Rammer1986}, which has only very recently been considered for problems of interfacial heat transfer~\cite{Wise2022}. Our computation, detailed in the SI Sec. 2.2, is closely analogous to the one carried out for quantum friction in ref.~\cite{Kavokine2022}. The theoretical framework can formally apply to fully non-equilibrium situations and take interactions into account to arbitrary order. However, to obtain a closed-form result, assume that the liquid and the solid internally equilibrated at temperatures $T_{\ell}$ and $T_{\rm e}$ respectively. Furthermore, we take electron-electron and electron-liquid Coulomb interactions into account at the Random Phase Approximation (RPA) level. With these assumptions, we obtain the electron-liquid energy transfer rate as
\begin{equation}
\mathcal{Q}_{\rm Q}  = \frac{1}{ 2\pi^3} \int \d \q  \int_0^{+\infty} \d \omega \, \hbar \omega [n_{\rm B}(\omega,T_{\rm e}) - n_{\rm B} (\omega,T_{\ell})] \frac{\Im{g_{\rm e}(\q,\omega)}\Im{g_{\rm \ell}(\q,\omega)}}{| 1 - g_{\rm e}(\q,\omega) g_{\rm \ell}(\q,\omega)|^2},
\label{result}
\end{equation}
Here, $n_{\rm B} (\omega, T) = 1/(e^{\hbar \omega/T} - 1)$ is the Bose distribution and the $g_{\rm e,\ell}$ are surface response functions of the solid and the liquid, respectively. These are analogues of the dielectric function for semi-infinite media, whose precise definition is given in the SI, Sec. 2.3. For the liquids under consideration, it will be sufficient to use the long-wavelength-limit expression of the surface response function: 
\begin{equation}
g_{\rm \ell} (q \to 0,\omega) = \frac{\epsilon_{\ell} (\omega) - 1 }{\epsilon_{\ell} (\omega) + 1}, 
\label{g_liq}
\end{equation}
where $\epsilon_{\ell} (\omega)$ is the liquid's bulk dielectric function. For two-dimensional graphene, we show in the SI (Sec. 2.3) that the surface response function can be expressed as 
\begin{equation}
g_{\rm e} (q, \omega) = - \frac{e^2}{2 \epsilon_0 q} \chi (q, \omega), 
\label{g_gra}
\end{equation}
where $\chi(q,\omega)$ is graphene's charge susceptibility. 

The result in Eq.~\eqref{result} has been derived for two solids separated by a vacuum gap in the framework of fluctuation-induced electromagnetic phenomena~\cite{Pendry1999,Volokitin2001,Volokitin2007}; our non-equilibrium framework, however, is better suited to the solid-liquid system under consideration. {We note that Eq.~\eqref{result} takes the form of a Landauer formula for the transport of bosonic quasiparticles -- elementary excitations of the solid's and the liquid's charge fluctuations modes~\cite{Biehs2021}. It involves indeed the difference of Bose distribution functions between the solid and the liquid, and the product of surface response functions plays the role of a transmission coefficient for the quasiparticles. One may count either the energy or the momentum transported by the quasiparticles: the former corresponds to near-field heat transfer, the latter to quantum friction. This quasiparticle picture thus makes explicit the fundamental connection between the two processes.}

\begin{figure}
\centering
\includegraphics{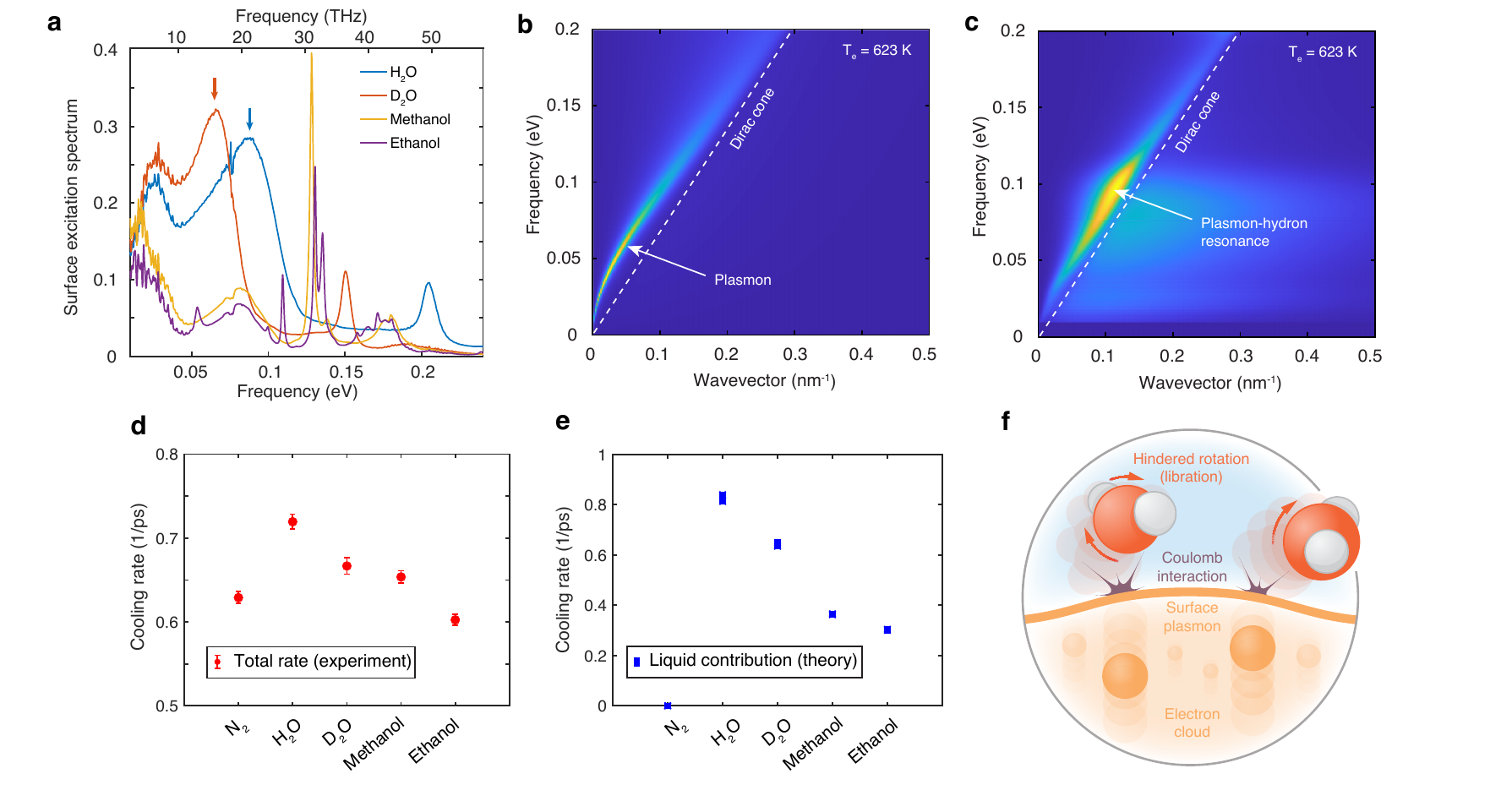}
\caption{\textbf{Mechanism of electron-liquid heat transfer}. \textbf{a}. Surface excitation spectra $\Im{g_{\ell} (\omega)}$ of the different liquids under study obtained according to Eq.~\eqref{g_liq} from the experimentally-measured bulk dielectric permittivities. The arrows indicate the libration modes of $\rm H_2O$ and $\rm D_2O$. \textbf{b}. Graphene surface excitation spectrum $\Im{ g_{\rm e} (q,\omega)}$, calculated at a chemical potential $\mu = 100~\rm meV$ and temperature $T_{\rm e} = 623 ~\rm K$. The main feature is the collective plasmon mode. \textbf{c}. Theoretical prediction for the graphene-water energy transfer rate resolved in frequency-wavevector space. The main contribution originates from a resonance between the graphene plasmon mode and the water libration mode. \textbf{d}. Experimentally-measured electron cooling rate in the presence of the various liquids. Error bars represent 95\% confidence intervals of the exponential fits to the temperature decay curves. \textbf{e}. Theoretical prediction for the liquid contribution to the electron cooling rate, reproducing the experimentally-observed trend in terms of the nature of the liquid. The symbol size in the vertical direction represents the variation in the theoretical prediction when the graphene chemical potential spans the range $\rm 100 - 180~meV$. \textbf{f}. Schematic of the water-mediated electron cooling mechanism inferred from the combination of theoretical and experimental results. The cooling proceeds through the Coulomb interaction between the graphene plasmon mode and the hindered molecular rotations (librations) in water. }
\end{figure}

\vskip0.5cm
\noindent{\bf \large Plasmon-hydron resonance} \\
{
The graphene electrons may relax either through direct interaction with the liquid, or through emission of optical phonons. The latter process has been well-studied, both theoretically and experimentally~\cite{Bistritzer2009,Pogna2021}. Our non-equilibrium formalism applies in principle to any electron-boson system: when applied to the electron-phonon system, it recovers the result for the energy transfer rate $\mathcal{Q}_{\rm ph}$ (from electrons to phonons) obtained in ref.~\cite{Pogna2021} (see SI, Sec. 2.3.2). Then, within a three-temperature model, where the electrons, liquid and phonons are assumed to be internally-equilibrated at temperatures $T_{\rm e}$, $T_{\ell}$ and $T_{\rm ph}$, respectively, we may determine the evolution of the electron temperature according to 
}
\begin{equation}
C(T_{\rm e}) \frac{\d T_{\rm e} (t)}{\d t} = - \mathcal{Q}_{\rm Q} (T_{\rm e}, T_{\ell}) - \mathcal{Q}_{\rm ph} (T_{\rm e}, T_{\rm ph}), 
\end{equation}
where $C(T_{\rm e})$ is the graphene electronic heat capacity at temperature $T_{\rm e}$. {We focus in the following on} the liquid contribution to the electron cooling rate, defined as $1/\tau = \mathcal{Q}_{\rm Q} (T_{\rm e}, T_{\ell}) / (C(T_{\rm e}) \times (T_{\rm e} - T_{\ell}))$, which may be compared with the experimental results. The quantitative evaluation of $\tau$ requires the surface response functions of graphene and of the various liquids. We compute the graphene surface response function according to Eq.~\eqref{g_gra} by numerical integration~\cite{Wunsch2006}, at the chemical potential determined for our samples by Raman spectroscopy (SI Sec. 1.4). For the liquids, we use the expression in Eq.~\eqref{g_liq}, with the bulk dielectric function determined by infrared absorption spectroscopy (Fig. 3a and SI Sec. 1.3). 

Our theoretical prediction for the various liquids' contribution to the electron cooling rate is shown in Fig. 3e. Quantitatively, we obtain cooling rates of the order of $1~\rm ps^{-1}$, in excellent agreement with the experimentally observed range (Fig. 3d) : our theory indicates that the quantum electron-liquid cooling is a sufficiently efficient process to compete with {the intrinsic phonon contribution, estimated at around $0.6~\rm ps^{-1}$ from the cooling rate in the absence of liquid}. Moreover, our theory reproduces the experimentally observed trend in cooling rates, with a significant liquid contribution arising only for water and heavy water; the dependence of the cooling rate on initial electron temperature is also well-reproduced (Fig. S7). Finally, the theory reproduces the isotope effect, that is, the slightly slower cooling observed with $\rm D_2O$ as compared to $\rm H_2O$. 

We may now exploit the theory to gain insight into the microscopic mechanism of the liquid-mediated cooling process. In Eq.~\eqref{result}, the difference of Bose distributions decreases exponentially at frequencies above $T_{\rm e} / \hbar \sim 100 ~\rm meV$. At frequencies below 100 meV, the graphene spectrum is dominated by a plasmon mode, that corresponds to the collective oscillation of electrons in the plane of the graphene layer~\cite{Wunsch2006} (Fig. 3b). In this same frequency range, water and heavy water have a high spectral density due to their libration mode, that corresponds to hindered molecular rotations~\cite{Carlson2020} (Fig. 3a). As a result, the energy transfer rate resolved in frequency-momentum space (the integrand in Eq.~\eqref{result}, plotted in Fig. 3c) has its main contribution from the spectral region where the two modes overlap. We conclude that the particularly efficient electron-water cooling is due to a resonance between the graphene plasmon mode and the water libration mode. This conclusion is further supported by the isotope effect. Indeed, the libration of the heavier $\rm D_2O$ is at slightly lower frequency than that of the lighter $\rm H_2O$, and a higher frequency mode makes a larger contribution to the cooling rate due to the factor $\hbar \omega$ in Eq.~\eqref{result}. {In the Landauer picture}, the quasiparticle {transport} rates are almost the same for the graphene-$\rm H_2O$ and graphene-$\rm D_2O$ systems, but in the case of $\rm H_2O$ each quasiparticle carries more energy. Overall, our experiments evidence a direct interaction between the graphene plasmon and water libration, as shown schematically in Fig. 3f. {We note that plasmons have been shown to play a role in the energy transfer between two graphene sheets~\cite{Ying2020}; however, to our knowledge, a plasmon-hydron interaction has so far not been suggested as a possible electron relaxation mechanism.}

\begin{figure}
\centering
\includegraphics{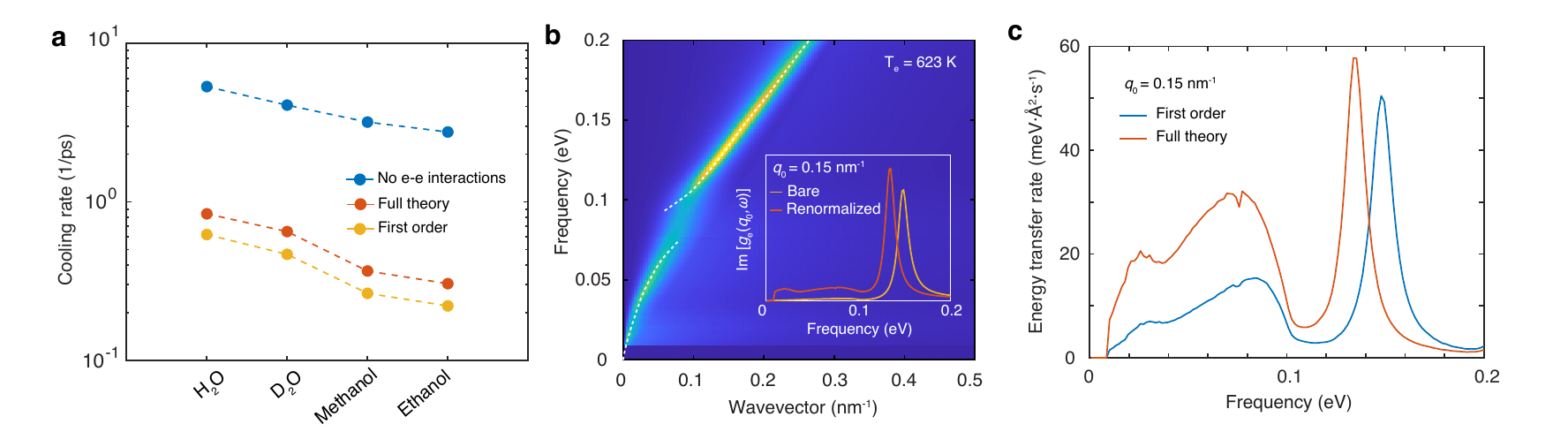}
\caption{\textbf{Strong plasmon-hydron coupling}. \textbf{a}. Theoretical prediction for the graphene electron cooling rate in contact with different liquids, within different treatments of interactions. The cooling rate is strongly overestimated if no electron-electron interactions are taken into account (blue symbols), and underestimated if the electron-liquid interactions are considered only to first order (orange symbols). \textbf{b}. Graphene surface excitation spectrum $\Im{ g_{\rm e} (q,\omega)}$, calculated at a chemical potential $\mu = 180~\rm meV$ and temperature $T_{\rm e} = 623 ~\rm K$, renormalized by the presence of water according to Eq.~\eqref{g_renorm}. The white dashed lines are guides to the eye showing the strongly-coupled plasmon-hydron mode. Inset: bare and renormalized graphene spectra at fixed wavevector $q_0 = 0.15 ~\rm nm^{-1}$. \textbf{c}. Comparison between the spectrally resolved energy transfer rates obtained to first order and to arbitrary order in the solid-liquid interaction. Higher-order effects enhance the energy transfer rate at low frequencies.}
\end{figure}

\vskip0.5cm
\noindent{\bf \large Interactions and strong coupling} \\
The combination of theory and experiment allows us to identify the key physical ingredients that are required to account for energy transfer at the water-graphene interface. First, our results reveal that electron-electron interactions are crucial, since they produce the plasmon mode that is instrumental to the energy transfer mechanism. Indeed, applying our theory to non-interacting graphene would result in a strongly overestimated liquid contribution to the cooling rate (Fig. 4a). This precludes single-particle Boltzmann approaches -- such as those that have been used for the electron-phonon interaction in graphene~\cite{Bistritzer2009,Pogna2021} -- for accurately describing the water-graphene interaction. 

Furthermore, the detailed examination of our theoretical result reveals that the efficiency of the electron-water cooling is enhanced by the formation of a strongly-coupled plasmon-hydron mode. Indeed, the result in Eq.~\eqref{result} involves bare surface response functions, without any renormalization due to the presence of the other medium. However, the denominator $|1 - g_{\rm e} g_{\rm \ell}|^2$ accounts for solid-liquid interactions to arbitrary order (at the RPA level) and contains the signature of any potential strong coupling effects. We find that these effects are indeed important, as removing the denominator in Eq.~\eqref{result} (that is, treating the electron-liquid interactions only to first order) results in under-estimation of the liquid-mediated cooling rate by about 30\% (Fig. 4a) . In order to gain physical insight into the nature of these higher order effects, we may compute the graphene surface response function renormalized by the presence of water, which is given by (see SI Sec. 2.3)
\begin{equation}
\tilde g_{\rm e} (q,\omega) = \frac{g_{\rm e}(q,\omega)}{1-g_{\rm e}(q,\omega) g_{\ell} (q,\omega)}. 
\label{g_renorm}
\end{equation}
The renormalized surface excitation spectrum $\mathrm{Im} \, [ \tilde g_{\rm e}(q,\omega) ] $ is plotted in Fig. 4b, for a chemical potential $\mu = 180~\rm meV$. We observe that the graphene plasmon now splits into two modes, which are both a mixture of the the bare plasmon and water libration. These are in fact analogous to the coupled plasmon-phonon modes that have been predicted~\cite{Hwang2010} and measured~\cite{Dai2015, Koch2016} for graphene on a polar substrate. It can be seen in the inset of Fig. 4b that coupling to the water modes also increases the spectral density at low frequencies (below the plasmon peak), compared to the bare graphene response function. This is in fact the higher-order effect that is mainly responsible for the enhancement of the electron cooling rate. As shown in Fig. 4c, taking into account solid-liquid interactions to arbitrary order mainly enhances the contribution of low frequencies to the energy transfer.

\vskip0.5cm
\noindent{\bf \large Conclusions} \\
We have carried out ultrafast measurements of electron relaxation in graphene, revealing signatures of direct energy transfer between the graphene electrons and the surrounding liquid. These results speak to the importance of electronic degrees of freedom in the dynamics of solid-liquid interfaces, particularly interfaces between water and carbon-based materials. Despite conventional theories and simulations that describe the interface in terms of atomic-scale Lennard-Jones potentials~\cite{Gutierrez2022, Herrero2022}, or with electronic degrees of freedom in the Born-Oppenheimer approximation~\cite{Tocci2014,Tocci2020}, here we demonstrate experimentally that the dynamics of the water-graphene interface need to be considered at the level of collective modes in the terahertz frequency range. In particular, our semi-quantitative theoretical analysis attributes the observed cooling dynamics to the strong coupling between the graphene plasmon and water libration modes. 

The experimental observation of such a collective mode interaction supports the proposed mechanism for quantum friction at the water-carbon interface, which is precisely based on momentum transfer between collective modes~\cite{Kavokine2022}. The near-quantitative agreement between the experiment and theory obtained for energy transfer suggests that a similar agreement should be achieved for momentum transfer. {The water-graphene quantum friction force is small if the graphene electrons are at rest, but becomes important if they are driven at a high velocity by a phonon wind or an applied voltage~\cite{Coquinot2022}. The quantum-friction-based driving of water flows by graphene electronic currents appears as a promising avenue in light of our findings. The electric circuit configuration would furthermore allow for noise thermometry~\cite{Yang2018,Baudin2020} to be used as a supplementary probe of the electron relaxation mechanisms.} 

Our results provide yet another example of the water-carbon interface outperforming other solid-liquid systems~\cite{Bocquet2020}. Indeed, the electronic contribution to the graphene-water thermal boundary conductance is as high as $\lambda = 0.25 ~\rm MW \cdot m^{-2} \cdot K^{-1}$, exceeding the value obtained with the other investigated liquids by at least a factor of 2. This even exceeds the thermal boundary conductance obtained for the graphene-hBN interface, at which particularly fast "super-Planckian" energy transfer was observed~\cite{Principi2017,Tielrooij2018}. Our investigation thus suggests that the density of modes in the terahertz frequency range is a key determinant for the thermal conductivity of graphene-containing composite materials.

\vskip0.5cm
\noindent{\bf \large Acknowledgements} \\
We acknowledge financial support from the MaxWater initiative of the Max Planck Society. We thank Xiaoyu Jia and Hai Wang for carrying out preliminary experiments, Maksim Grechko and Detlev-Walter Scholdei for assisting with the FTIR measurements, Marc-Jan van Zadel and Florian Gericke for constructing the sample holder. X.Y. is grateful for support from the China Scholarship Council. K.J.T. acknowledges funding from the European Union's Horizon 2020 research and innovation program under Grant Agreement No. 804349 (ERC StG CUHL), RYC fellowship No. RYC-2017-22330, and IAE project PID2019-111673GB-I00. A.P. acknowledges support from the European Commission under the EU Horizon 2020 MSCA-RISE-2019 programme (project 873028 HYDROTRONICS) and from the Leverhulme Trust under the grant RPG-2019-363. N.K. acknowledges support from a Humboldt fellowship. The Flatiron Institute is a division of the Simons Foundation. We thank Lucy Reading-Ikkanda (Simons Foundation) for help with figure preparation. 

\vskip0.5cm
\noindent{\bf \large {Author contributions}} \\
{M.B., K.-J.T. and N.K. conceived the project. X.Y. carried out the experiments and analyzed the data. N.K. developed the theoretical model and wrote the paper. All authors discussed the results and commented on the manuscript.}

\vskip0.5cm
\noindent{\bf \large {Competing interests}} \\
{The authors declare no competing interests.}

\vskip0.5cm
\noindent{\bf \large {Data availability}} \\
{The experimental data supporting the findings is available on Zenodo. DOI: 10.5281/zenodo.7738429}

\end{document}


\begin{titlepage}

\tableofcontents
\end{titlepage}

\section{Experimental methods}

\subsection{Sample preparation}

CVD-grown graphene samples supported on 1 mm-thick copper substrates were purchased from Grolltex Inc. The MilliQ water ($18.2~\rm M\Omega\cdot  cm$) was used as obtained from the machine. Cellulose acetate butyrate (CAB, average Mn $\sim 12000$, Sigma-Aldrich), ammonium persulfate (APS, ACS reagent, $\geq 98\%$, Honeywell Fluka$^{\text{\sc TM}}$) are used as received. CAB was dissolved in ethyl acetate (Sigma-Aldrich), producing a 30 mg/mL solution. APS was dissolved in MilliQ water to prepare 1~M and 0.1~M solutions. The detachable fused silica flow cell was ordered from FireflySci, Inc. The flow cell was cleaned by sonication in a hot acetone and ethanol baths for 10 minutes each before using.

We transferred graphene onto the front substrate of the flow cell following a wet transfer procedure~\cite{Yogeswaran2018,Burwell2015}. First, we spin-coated graphene samples with CAB at 4000 rpm and baked them at $180\rm {}^{\circ} C$ for 3 minutes. Then, to remove unnecessary graphene on the backside of copper substrates, the CAB-coated graphene samples were immersed into a 1~M solution of APS for 10 minutes and subsequently rinsed with MilliQ water five times. The copper substrates were then fully etched by 0.1~M APS solution for 2 hours, followed by a five times rinse with MilliQ water to remove the attached ions. Then, the floating CAB-graphene monolayers were "fished" onto the flow cell, and the CAB coating was removed by soaking in acetone for 2 hours and in isopropanol for one hour. 

\subsection{OPTP measurements}

\begin{figure}
\centering
\includegraphics[width=12cm]{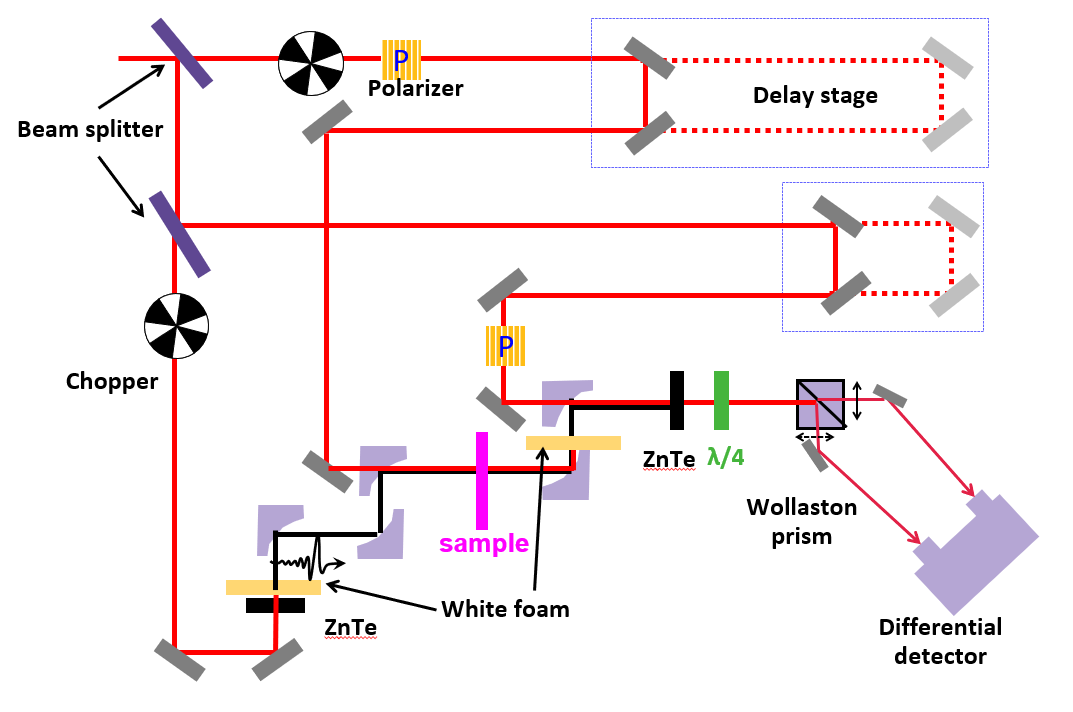}
\caption{Schematic of the OPTP setup.}
\end{figure}

We probed electron relaxation in graphene using optical pump - terahertz probe (OPTP) spectroscopy. A schematic of the OPTP setup is shown in Fig. S1. 
The fundamental laser output was generated by a regenerative Ti:sapphire amplifier system, which produces 5 W, 50 fs pulses at a repetition rate of 1 kHz and a central wavelength of 800 nm. The generated pulses were then split into three branches for THz generation, sampling, and optical excitation. A single-cycle THz pulse of $\sim 1~\rm ps$ duration was generated by pumping a 1 mm thick (110) ZnTe crystal with the 800 nm fundamental pulses via optical rectification. 

We photoexcited graphene to generate hot carriers by using 800 nm pulses with a diameter of 5 mm to ensure a homogeneously photoexcited region. The transmitted THz wave was then recollimated and focused onto a ZnTe detection crystal together with an 800 nm sampling beam, where the THz electrical field waveform was detected using the electro-optic sampling method~\cite{Ulbricht2011}. The THz pulse induces birefringence in the ZnTe detection crystal, and the polarization of the sampling beam is thus changed. After passing through a quarter-wave plate, the sampling beam changes from perfectly circular to slightly elliptical shape. The $s$ and $p$ components of this elliptically polarized pulse are separated by a Wollaston prism, and the difference of these two components is detected by a balance diode. The signal is collected by a lock-in amplifier that is phase-locked to an optical chopper that modulates either the THz generation beam or the pump beam at a frequency of 500 Hz. The ultrafast time evolution of the peak intensity of the THz field is tracked by varying the time delay between optical pump and THz probe~\cite{Ulbricht2011, Lee2009}. The setup was purged with dry nitrogen during the measurement to avoid the absorption of water vapor. 

The raw data consists in time traces of the pump-induced transmission change at the peak of the THz waveform ($\Delta E$), normalized by the peak value of the THz transmission without excitation ($E$) (Fig. S2a). Assuming that a fraction $\gamma = 1.6\%$ of the pump pulse energy is absorbed by the graphene electrons~\cite{Fu2021}, the maximum electron temperature reached after photoexcitation can be related to the pump laser fluence $F$ according to $\gamma F = C(T_{\rm e}) T_{\rm e}$, where $C(T_{\rm e})$ is the graphene heat capacity at temperature $T_{\rm e}$. In the limit where the graphene Fermi energy $\mu$ is larger than $k_{\rm B}T_{\rm e}$ (as relevant for our samples), we may use the approximate expression~\cite{Shi2014,Tielrooij2013,Lui2010}
\begin{equation}
C(T_{\rm e}) = \alpha T_{\rm e}, ~~~\text{with} ~~~ \alpha = \frac{2 \pi}{3} \frac{k_{\rm B}^2 \mu}{(\hbar v_{\rm F})^2}, 
\end{equation}
where $v_{\rm F}$ is graphene's constant Fermi velocity. Then, 
\begin{equation}
T_{\rm e} = T_0 \left( 1 + \frac{2 \gamma F}{\alpha T_0^2} \right)^{1/2}, 
\end{equation}
where $T_0$ is ambient temperature. The peak value of $\Delta E/ E$ after photoexcitation increases with laser fluence. Upon rescaling, we find that the plots of $\Delta E/ E$ vs. $F$ and $T_{\rm e}$ vs. $F$ collapse upon each other (Fig. S2b), so that we may consider that $\Delta E/E$ is proportional to the electron temperature within the range of temperatures probed in the experiment, as shown explicitly in Fig. S2c. 

\begin{figure}
\centering
\includegraphics[width=\textwidth]{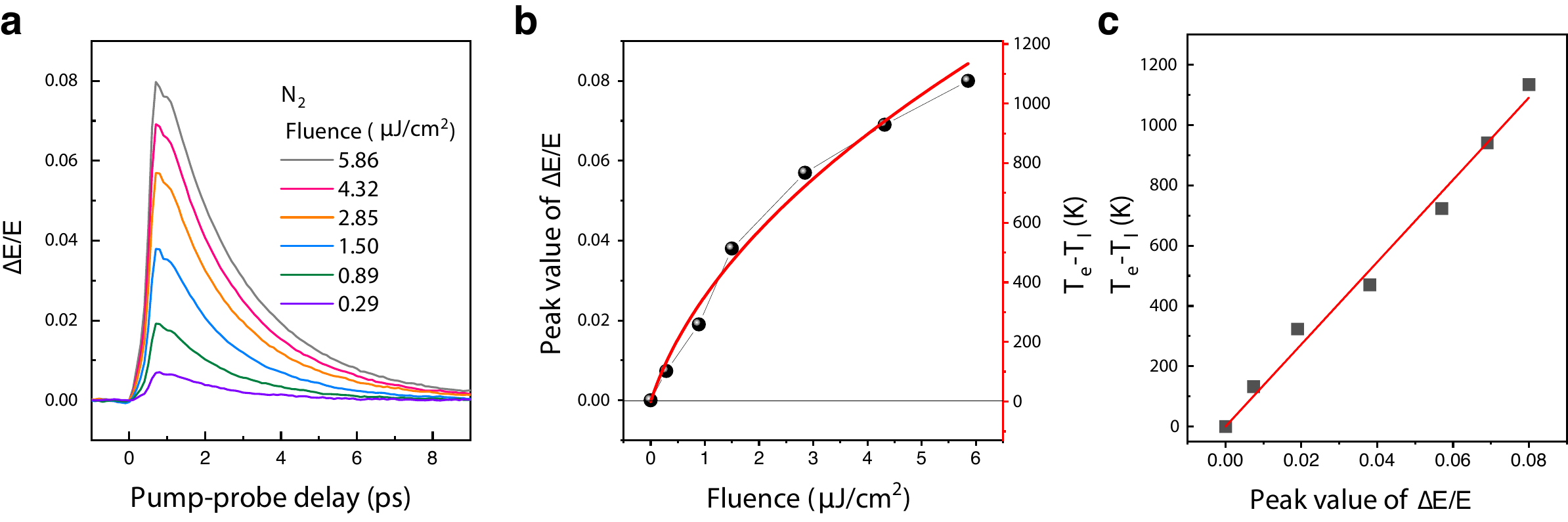}
\caption{\textbf{Electron temperature of the pumped graphene layer}. \textbf{a}.The OPTP traces of graphene in a nitrogen atmosphere with various excitation fluences. \textbf{b}. Peak value of $\Delta E / E$ as a function of laser fluence and corresponding electron temperature. \textbf{c}. Increase in electron temperature ($T_{\rm e}$) with respect to ambient temperature $T_{\ell}$ as a function of $\Delta E/ E$: a linear relation is observed.}
\end{figure}

The thickness of the liquid layer was set to $50~\rm \mu m$ by the geometry of the flow cell. The liquids were exchanged using a syringe and the spectroscopic measurement was always carried out at the same spot of the graphene sample. To exclude the effect of beam dispersion in the different liquids on the results, we repeated the measurement with different water layer thickness and using different Teflon spacers between two fused silica windows (Fig. S3 a and b). {We further checked that there was no THz signal from water in the absence of graphene (Fig. S3c). }

\begin{figure}
\centering
\includegraphics[width=\textwidth]{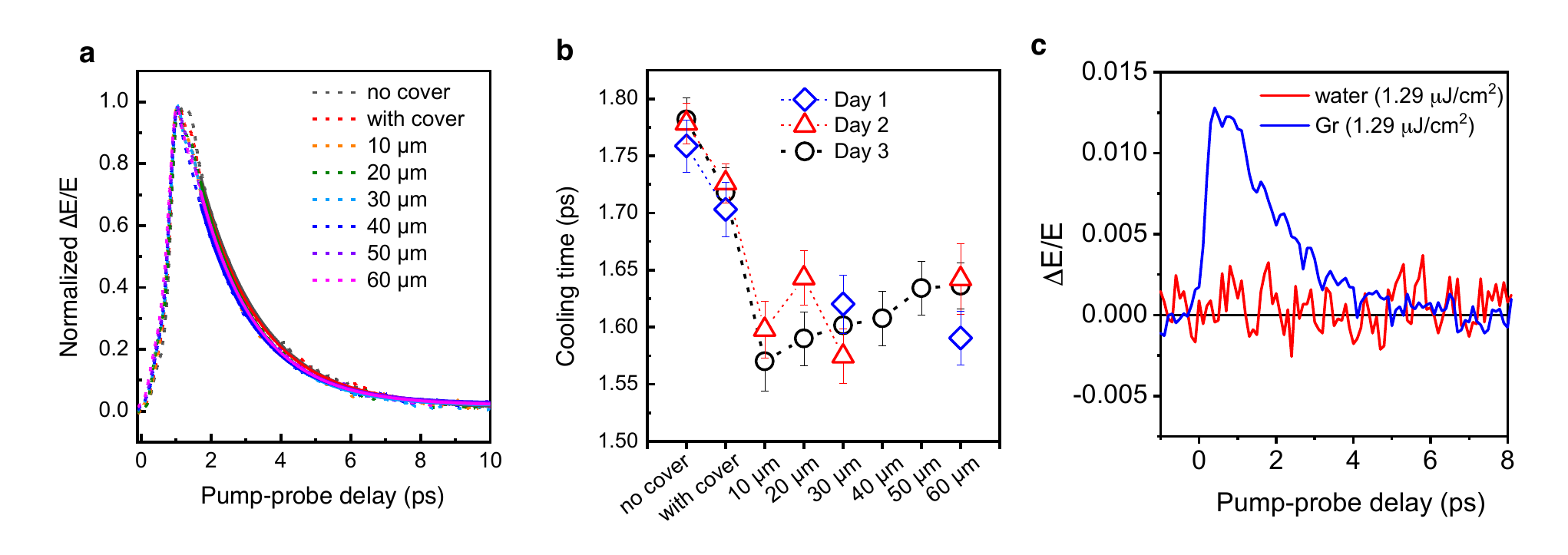}
\caption{\textbf{Control experiments}. \textbf{a}. The OPTP traces of graphene with varying water layer thickness. \textbf{b}. Cooling times obtained by exponential fitting of the data in panel a. The error bars represent 95\% confidence intervals of the exponential fits. \textbf{c}. {Comparison of the THz signals after photoexcitation, with and without graphene. Water in the absence of graphene shows no measurable THz response.} }
\end{figure}

\subsection{FTIR measurements}

\begin{figure}
\centering
\includegraphics[scale=0.7]{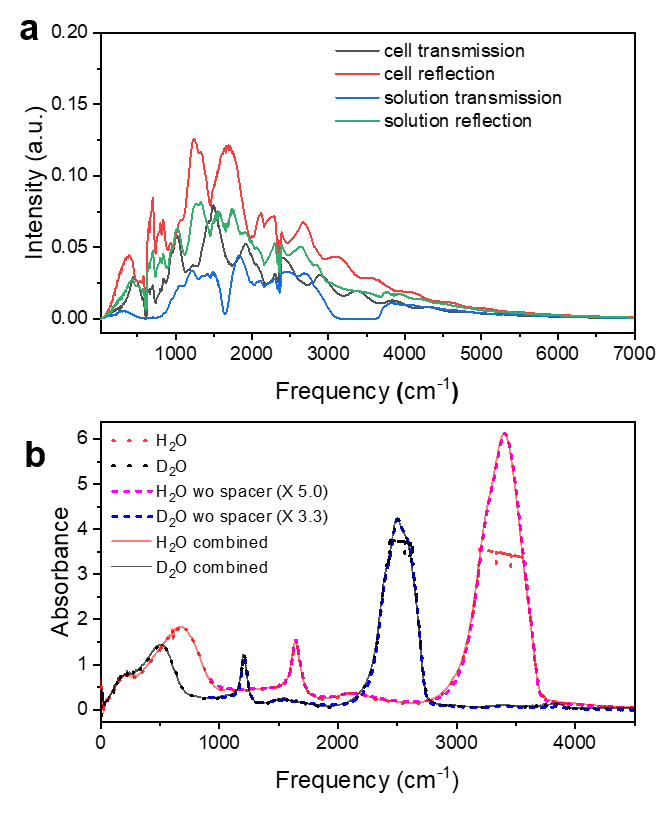}
\caption{\textbf{FTIR data analysis}. \textbf{a}. Raw intensity data of empty cell and water-filled cell. \textbf{b}. Absorbance of  $\rm H_{2}O$ and $\rm D_{2}O$, measured with spacer and without spacer (the latter is rescaled to overlap with the former).}
\end{figure}

\begin{figure}
\centering
\includegraphics[scale=0.8]{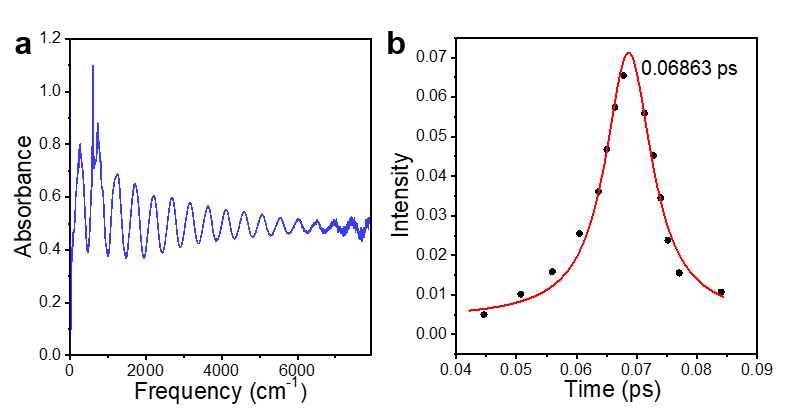}
\caption{\textbf{Determination of the cell thickness}. \textbf{a}. Absorbance of empty cell. \textbf{b}. Fourier transform of the data in panel a.}
\end{figure}

We measured the dielectric functions of water, heavy water, ethanol and methanol using Fourier-transform infrared (FTIR) spectroscopy. We measured the transmitted and reflected infrared intensities both for an empty cell $(I_{\rm t, cell}, I_{\rm r, cell})$ and for a cell filled with liquid $(I_{\rm t, liquid}, I_{\rm r, liquid})$ thanks to an A510/Q-T Reflectance and Transmittance accessory placed in a commercial VERTEX 70 FTIR spectrometer (Fig. S4a). In order to avoid disassembling the cell when changing liquids, we carried out the measurements inside a flow cell, made out of two-silicon wafers separated by a $10 ~\rm \mu m$ Teflon spacer. 

 We calculated the absorbance $A(\omega)$ according to 
\begin{equation}
A(\omega) = - \log_{10} \left( \frac{I_{\rm t, solution} (\omega) }{I_{\rm t, cell} (\omega) + I_{\rm r, cell} (\omega) - I_{\rm r, solution} (\omega) } \right). 
\end{equation}
The $\rm H_{2}O$ and $\rm D_{2}O$ show saturated absorption in the range of 3100-3600 and 2200-2700 $\rm cm^{-1}$, respectively. We obtained the data in this frequency range by measuring the spectra without any spacer between two $\rm CaF_2$ windows and then rescaled the spectra to overlap with the data with spacer (Fig. S4b). 
The imaginary part $k(\omega)$ of the refractive index is related to the absorbance by 
\begin{equation}
k(\omega) = A(\omega) \frac{\ln (10)}{4 \pi \omega \ell},
\end{equation}
where $\ell$ is the sample thickness. To accurately determine the thickness of the cell, we calculate the absorbance of the empty cell without correction for multiple reflections, 
\begin{equation}
A_2(\omega) = - \log_{10} \left( \frac{I_{\rm t, cell} (\omega) }{I_{\rm t, lamp} (\omega)} \right). 
\end{equation}
where $I_{\rm t, lamp} (\omega)$ is the intensity of the lamp of the FTIR source (Fig. S5a). Fourier transformation of this spectrum yields a peak at the time $\Delta t$ that light takes to travel twice through the cell (Fig. S5b), so that $\ell = c \Delta t/2 =10.29~\rm \mu m$. 
We then obtained the real part of the refractive index through a numerical Kramers-Kr\"{o}nig transformation: 
\begin{equation}
n(\omega) = n_{\infty} + \frac{2}{\pi} \int_0^{\infty} \d \omega' \,  \frac{k(\omega') }{\omega' - \omega}, 
\end{equation}
where $n_{\infty}$ is the refractive index in the high frequency limit, which is obtained by the ATAGO Digital Handheld Refractometer: PAL-RI. The measured values for $\rm H_{2}O$, $\rm D_{2}O$, methanol, ethanol and isopropanol are 1.333, 1.3291, 1.3285, 1.3604, and 1.3706 respectively.

We then obtain the dielectric function $\epsilon(\omega) = \epsilon'(\omega) + i \epsilon''(\omega) $ according to 
\begin{equation}
\left\{
\begin{array}{l}
\epsilon'(\omega) = n(\omega)^2 - k(\omega)^2 \\
 \epsilon''(\omega) = 2 n(\omega) k(\omega) \\
\end{array}
\right. .
\end{equation}

\subsection{Raman measurements}

\begin{figure}
\centering
\includegraphics{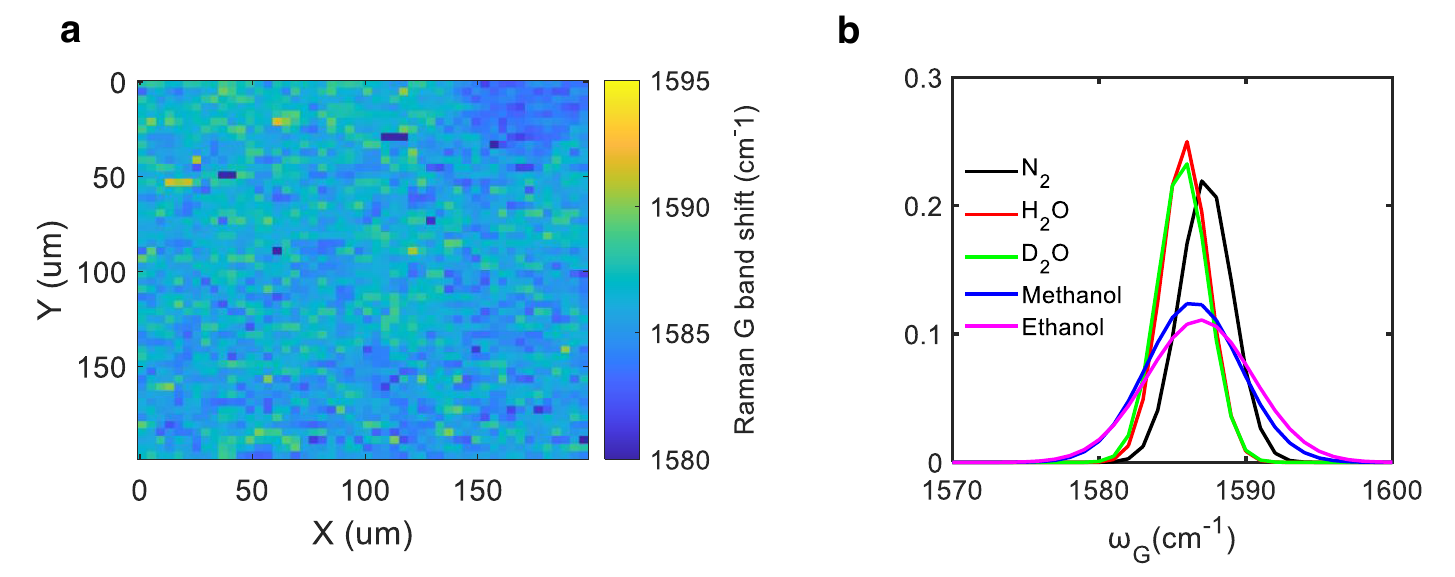}
\caption{\textbf{Raman characterization of graphene sample}. \textbf{a}. Spatial map of Raman G band frequency for graphene sample in air. \textbf{b}. Distribution of the Raman G band frequency with different liquids placed on the graphene surface.}
\end{figure}

We estimated the Fermi level $\mu$ in our liquid-covered graphene samples from the Raman G-band frequency, according to the empirical equation~\cite{Shi2014}
\begin{equation}
|\mu| (\mathrm{eV}) = \frac{\omega_{\rm G} - 1580 ~\rm cm^{-1}}{42 ~\rm cm^{-1}}.
\end{equation}
An example of a spatial map of the Raman G-band frequency is shown in Fig. S6a. The frequency shows spatial inhomogeneities on the $\mu$m scale with an amplitude around $10~\rm cm^{-1}$. The corresponding distributions are shown in Fig. S6b. The average G-band frequency is essentially independent of the nature of the liquid, which excludes a change in charge carrier density as a possible mechanism for the liquid effect on the electron cooling rate. To take into account the broadness of the distribution, in the theoretical analysis we considered chemical potentials in the range $\mu = 100 - 180~\rm meV$. The theoretical prediction is independent of the electron or hole nature of the charge carriers. 

\section{Theoretical methods}

In this section, we develop a description of energy transfer between the Dirac fermion charge carriers in graphene and a liquid, treated as a bosonic bath, within the non-equilibrium Keldysh framework of perturbation theory. For the sake of completeness, and in order to show consistency with previous theoretical approaches, we apply the same description to energy transfer between the graphene electrons and its optical phonon modes, showing that our formalism recovers the results that were previously obtained within a Boltzmann equation approach~\cite{Pogna2021}. 

\emph{We use SI units throughout the text. We adopt the following convention for the $n$-dimensional Fourier transform:}
\begin{equation}
\tilde f (\q) = \int_{-\infty}^{+\infty} \d^n \rr\, f(\rr) e^{-i \q \cdot \rr} ~~~~ \mathrm{and} ~~~~ f(\rr) =  \frac{1}{(2\pi)^n}\int_{-\infty}^{+\infty} \d^n \q\, \tilde f(\q) e^{i \q \cdot \rr}.
\end{equation} 

\subsection{Interaction Hamiltonian}

\subsubsection{Electron-hydron interaction}

\emph{In this section, $\rr$ represents a vector in 3D space, and $\boldsymbol{\rho}$ a vector in 2D space}. The charge fluctuations of the liquid in the $z>0$ half-space couple to the graphene electrons via the Coulomb potential $V$. In real space, the corresponding Hamiltonian is 
\begin{equation}
H_{\rm ew}(t) = \int \d \rr \d \rr' n_w(\rr,t) V(\rr - \rr') n_e(\rr,t), 
\end{equation}
where $n_w$ and $n_e$ are the liquid and graphene instantaneous charge density, respectively. Let $c^{\dagger}_{\k,\nu}, c_{\k,\nu}$ be the Dirac fermion creation and annihilation operators in the chiral basis ($\nu = \pm 1$). A 2D Fourier transformation then yields 
\begin{equation}
H_{\rm int} =  \int \frac{\d \q}{(2\pi)^2} \frac{e^2}{2\epsilon_0 q} n_{\rm s}(q,t) \sum_{\k,\nu,\nu'} \langle \k + \q, \nu | e^{i \q \brho} e^{qz} |\k,\nu' \rangle c^{\dagger}_{\k+\q,\nu}(t) c_{\k,\nu'}(t),
\end{equation}
with 
\begin{equation}
n_{\rm s}(q) = \int \d \brho \int_0^{+\infty} \d z \, e^{-i \q \brho} e^{-qz} n_{\rm w}(\brho,z,t). 
\label{ns}
\end{equation}
As long as we consider wavevectors $\q$ such that $q^{-1}$ is large compared to the extension of the carbon $p_z$ orbitals perpendicular to the graphene plane, we may approximate
\begin{equation}
 | \langle \k + \q, \nu | e^{i \q \brho} e^{qz} |\k,\nu' \rangle|^2 \approx |\langle \mathbf{k} + \mathbf{q},\nu | e^{i \q  \brho}| \mathbf{k} ,\nu'\rangle|^2 = \frac{1}{2} \left( 1+ \nu \nu' \cos (\phi_{\k+\q}-\phi_{\q}) \right). 
\end{equation}

\subsubsection{Electron-phonon interaction}
Let $d^{\dagger}_{\q,\alpha}, d_{\q,\alpha}$ be the creation and annihilation operators of phonons in the mode $\alpha$ with frequency $\omega_{\alpha}$. The non-interacting electron-phonon system's Hamiltonian is 
 \begin{equation}
 H_0 = \sum_{\k,\nu} E_{\k,\nu} c^{\dagger}_{\k,\nu} c_{\k,\nu} + \sum_{\q,\alpha} \hbar \omega_{\alpha} d^{\dagger}_{\q,\alpha} d_{\q,\alpha}, 
 \end{equation}
where $E_{\k,\nu}$ are the band energies, and $\sum_{\k} \equiv (1/\mathcal{A}_{\rm BZ}) \int_{\rm BZ} \d \k$ ($\mathcal{A}_{\rm BZ}$ is the area of the 2D Brillouin zone). The electron-phonon interaction Hamiltonian has the general form~\cite{CastroNeto2007}
\begin{equation}
H_{\rm ep} = \sum_{\alpha} \int_{\rm BZ} \frac{\d \q}{(2\pi)^2} \sum_{\k,\nu,\nu'} g_{\alpha,\k,\k+\q}^{\nu \nu'} c^{\dagger}_{\k +\q} c_{\k} (d^{\dagger}_{\q,\alpha} + d_{-\q,\alpha}),  
\end{equation}
Following~\cite{Pogna2021}, we consider the $\Gamma$ point LO and TO phonons that scatter electrons within one valley, and the K, K' point LO phonons that scatter electrons between valleys. The electron-phonon matrix elements read
\begin{equation}
| g_{\Gamma,\k,\k+\q}^{\nu \nu'} |^2 = g_{\Gamma}^2 (1 \pm \nu \nu' \cos(\phi_{\k} + \phi_{\k + \q} - 2 \phi_{\q}) ),  
\end{equation}
where the $+$  $(-)$ sign is for LO (TO) phonons; and 
\begin{equation}
| g_{\Gamma,\k,\k+\q}^{\nu \nu'} |^2 = g_{\rm K}^2 (1 \mp \nu \nu' \cos(\phi_{\k} - \phi_{\k + \q} ) ),
\end{equation}
where the $-$ $(+)$ sign corresponds to scattering from K to K' (from K' to K); here, $\phi_{\v}$ is the polar angle of the vector $\v$. The values of the coupling constants are $g_{\Gamma} = 0.55~ \rm eV \cdot \angstrom$ and $g_{\rm K} = 0.85~\rm eV \cdot \angstrom$, according to GW calculations~\cite{Sohier2014}.

\subsubsection{General form} 

We find that for both types of interactions the Hamiltonian has the general form 
\begin{equation}
H_{\rm eb} = \int \frac{\d \q}{(2\pi)^2} n_{\q}(t) \varphi_{\q} (t), 
\end{equation}
where $n_{\q}$ is an electronic two-particle operator and $\varphi_{\q}$ is a free bosonic field. In the electron-phonon case, we define 
\begin{equation}
n_{\q} = \sum_{\k,\nu,\nu'} \frac{g_{\alpha,\k,\k+\q}^{\nu \nu'}}{\sqrt{\hbar\omega_{\alpha}}} c^{\dagger}_{\k +\q,\nu} c_{\k,\nu'} ~~~ \text{and} ~~~ \varphi_{\q} = \sqrt{\hbar \omega_{\alpha}} (d^{\dagger}_{\q,\alpha} + d_{-\q,\alpha}); 
\end{equation}
in the electron-hydron case 
\begin{equation}
n_{\q} = \sqrt{V_q}\sum_{\k,\nu,\nu'} \langle \mathbf{k} + \mathbf{q},\nu | e^{i \q  \brho}| \mathbf{k} ,\nu'\rangle c^{\dagger}_{\k +\q,\nu} c_{\k,\nu'} ~~~ \text{and} ~~~ \varphi_{\q} = \sqrt{V_q}n_{\rm s}(\q),
\label{defphiw}
\end{equation}
where $V_q \equiv e^2/(2\epsilon_0 q)$ is the 2D Fourier-transformed Coulomb potential. With these definitions, both $n_{\q}$ and $\phi_{\q}$ have dimensionless correlation functions in frequency space. 

\subsection{General theory of electron-boson heat transfer}
\subsubsection{Non-equilibrium perturbation theory}

We consider an initial state of the electron-boson system where the electrons are at a temperature $T_{\rm e}$ and the bosons at a temperature $T_{\rm b}$. We wish to study the subsequent dynamics. In particular, we are interested in the heat flux per unit surface from the electrons to the bosons: 
\begin{equation}
\mathcal{Q}(t) = - \frac{1}{\mathcal{A}} \, \frac{\d}{\d t} \langle H_{\rm eb} (t) \rangle. 
\end{equation}
Since the system is under non-equilibrium conditions, this average value needs to be computed in the Keldysh framework. In particular, we may define the Keldysh component of the electron-boson correlation function: 
\begin{equation}
\chi_{\rm eb}^{K} (\q,t,t')  = - \frac{1}{\mathcal{A}} \frac{i}{\hbar}  \langle \{ n_{\q}(t), \varphi_{-\q}(t') \} \rangle. 
\end{equation}
Then, 
\begin{equation}
\mathcal{Q} (t) = - \frac{i\hbar}{2} \int \frac{\d \q}{(2\pi)^2} \frac{\d \chi_{\rm eb}^{K} (\q,t,t)}{\d t}. 
\label{defQ}
\end{equation}
Form this point on, the computation of the electron-boson correlation function follows the exact same steps as in the theory of quantum friction~\cite{Kavokine2022}, and we reproduce here only the main equations. Diagramatically, the correlation function satisfies the following Dyson equation: 
\begin{equation}
\centering
\includegraphics{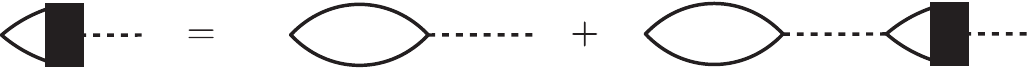}
\end{equation}
where the "bubble" represents the propagator of $n$ (denoted $\chi_e$), and the dashed line the propagator of $\varphi$ (denoted $\chi_b$). When made explicit in terms of the $R,A,K$ components, the Dyson equation becomes 
\begin{equation}
\left\{
\begin{array}{l}
\chi_{\rm eb}^K = \chi_{\rm e}^R \otimes \chi_{\rm b}^K + \chi_{\rm e}^K \otimes \chi_{\rm b}^A + \chi_{\rm e}^R \otimes \chi_{\rm b}^R \otimes \chi_{\rm eb}^K + (\chi_{\rm e}^R\otimes\chi_{\rm b}^K+\chi_{\rm e}^K\otimes \chi_{\rm b}^A)\otimes \chi_{\rm eb}^A\\
\\
\chi_{\rm eb}^{R,A} = \chi_{\rm e}^{R,A}\otimes \chi_{\rm b}^{R,A} + \chi_{\rm e}^{R,A} \otimes \chi_{\rm b}^{R,A} \otimes \chi_{\rm eb}^{R,A}
\end{array}
\right. ,
\label{dyson2}
\end{equation}
where $\otimes$ represents time convolution. While these equations are extremely general, they are impractical to manipulate analytically, unless a number of assumptions are made. In order to proceed, we will restrict ourselves to cooling dynamics that are slow enough for time-translation invariance to hold when it comes to determining the cooling rate. This assumption is expected to hold for small enough temperature differences, such that the cooling rate is approximately temperature-independent. We will further assume that, in line with experimental observations, that electron thermalization is much faster than electron-boson energy transfer, so that the electron and boson propagators may be considered as equilibrium propagators, satisfying the fluctuation-dissipation theorem: we work within a two-temperature model. We may then carry out Fourier transforms in time, so that Eq.~\eqref{defQ} becomes
\begin{equation}
\mathcal{Q}  =  \frac{1}{2} \int \frac{\d \q \d \omega}{(2\pi)^3} \hbar \omega \, \chi^K_{\rm eb} (\q,\omega). 
\label{defQ2}
\end{equation}
The convolutions in Eq.~\eqref{dyson2} become products in Fourier space. Before proceeding, it is convenient to flip the signs of all the correlation functions: we introduce, for all the labels, $g \equiv - \chi$. Then, after some algebra, we obtain an explicit expression for $\mathcal{Q}$: 
\begin{equation}
\mathcal{Q} = \frac{1}{ 2\pi^3} \int \d \q  \int_0^{+\infty} \d \omega \, \hbar \omega [n_{\rm B}(\omega,T_{\rm e}) - n_B (\omega,T_{\rm b})] \frac{\Im{g_{\rm e}^R(\q,\omega)}\Im{g_{\rm b}^R(\q,\omega)}}{| 1 - g_{\rm e}^R(\q,\omega) g_{\rm b}^R(\q,\omega)|^2},
\label{result}
\end{equation}
where $n_{\rm B}(\omega,T) \equiv 1/(e^{\hbar \omega/k_B T} - 1)$ is the Bose distribution at temperature $T$. We recover Eq.~(3) of the main text. 

\subsubsection{Cooling rate}
The cooling dynamics are governed by the equation 
\begin{equation}
\frac{\d \mathcal{E}(T_{\rm e})}{\d t} = - \mathcal{Q}(T_{\rm e},T_{\rm b}), 
\end{equation}
where $\mathcal{E}$ is the total energy per unit surface of the electronic system. We follow ref.~\cite{Pogna2021} in determining the electronic heat capacity (per unit surface) at constant density $C(T_{\rm e})$, such that $\d_t \mathcal{E} = C(T_{\rm e}) \d_t T_{\rm e}$. We may then define the instantaneous cooling rate 
\begin{equation}
\tau (T_{\rm e}, T_{\rm b}) = \frac{C(T_{\rm e}) (T_{\rm e}-T_{\rm b})}{\mathcal{Q}(T_{\rm e},T_{\rm b})}. 
\label{tau}
\end{equation}

\subsection{Application to the graphene-liquid system}

\subsubsection{Liquid-mediated cooling}

We first consider electron cooling through the electron-hydron coupling. Using eqs.~\eqref{ns} and \eqref{defphiw}, we find that 
\begin{equation}
g_{\rm b}^R(\q,t,t') = -\frac{1}{\mathcal{A}} V_q \int_0^{+\infty} \d z \d z' \, e^{-q(z+z')} \left[ -\frac{i}{\hbar} \theta (t-t') \langle [n_{\rm s}(\q,z,t),n_{\rm s}(-\q,z',t')] \rangle \right].
\end{equation}
This is the microscopic definition of the liquid's \emph{surface response function}. In the long wavelength limit, it can be expressed in terms of the liquid's bulk dielectric function $\epsilon(\omega)$~\cite{Kavokine2022}:
\begin{equation}
g_{\rm b}^R(\q,\omega) = \frac{\epsilon(\omega)-1}{\epsilon(\omega)+1}, 
\end{equation}
as stated in the main text. The electronic response function $g_{\rm e}^R(\q,\omega)$ simply amount to (minus) the density-density response function. Taking into account electron-electron interactions at the RPA level~\cite{Wunsch2006}, 
\begin{equation}
g_{\rm e}^R(q,\omega) = - \frac{V_q \chi^0_{\rm e} (q,\omega)}{1-V_q \chi^0_{\rm e}(q,\omega)}.
\end{equation}
The non-interacting response function $\chi^0_e$ is given by~\cite{Wunsch2006}
\begin{equation}
\chi^0_{\rm e} (q,\omega) = g_{\rm s} g_{\rm v} \int \frac{\d \k}{(2 \pi)^2} \sum_{\nu,\nu'} |\langle \mathbf{k} + \mathbf{q},\nu | e^{i \q  \brho}| \mathbf{k} ,\nu'\rangle|^2 \frac{n_{\rm F} (E_{\k}^{\nu},T_{\rm e}) - n_{\rm F} (E_{\k+\q}^{\nu'},T_{\rm e})}{E_{\k}^{\nu} - E_{\k+\q}^{\nu'} + \omega + i \delta}, 
\end{equation}
where $g_{\rm s} = g_{\rm v} = 2$ are the spin and valley degeneracies of graphene, respectively, $E_{\k}^{\nu} = \nu v_F k$ are the band energies in the Dirac fermion approximation, $n_{\rm F}(E,T) = 1/(e^{(E-\mu)/k_B T}+1)$ is the Fermi distribution at chemical potential $\mu$ and temperature $T$, and $\delta \to 0^+$. The integral is evaluated numerically at non-zero temperature. 

With all the above, we may compute theoretical predictions for the liquid-mediated cooling rate by numerical integration according to Eq.~\eqref{result}. We considered a graphene chemical potential $\mu$ in the range $100 - 180~\rm meV$ (see section 1.4) and an electron temperature $T_{\rm e} = 623~\rm K$, corresponding to the lowest pump laser fluence.  Our model is further able to reproduce the dependence of the electron cooling time on $T_{\rm e}$, as shown in Fig. S7. 

\begin{figure}
\centering
\includegraphics{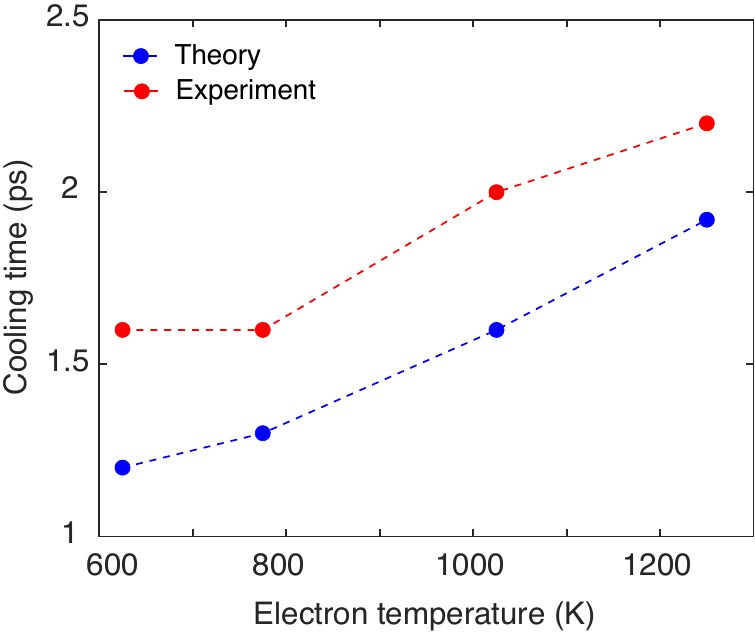}
\caption{Dependence of water-mediated cooling time on initial electron temperature. The red dots are experimental data for graphene in contact with water and the red dots correspond to the prediction of Eq.~\eqref{tau} (with $\mu = 180~\rm meV$).}
\end{figure}

We note that Eq.~\eqref{result} involves bare surface response functions, that contain no effect of the presence of the neighboring medium, at least at the RPA level. Nevertheless, the physical response function of graphene in the presence of water undergoes RPA renormalization according to 
\begin{equation}
\centering
\includegraphics{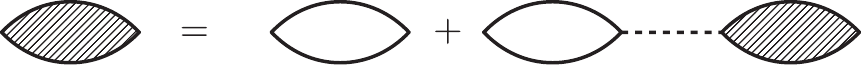}
\end{equation}
In this diagrammatic equation, when the propagators are interpreted as surface response functions, the vertices reduce to unity, so that we obtain the renormalized graphene response function $\tilde g_{\rm e}$ as 
\begin{equation}
\tilde g_{\rm e} (q,\omega) = \frac{g_{\rm e}(q,\omega)}{1-g_{\rm e}(q,\omega) g_{\rm b} (q,\omega)},
\label{g_renorm}
\end{equation}
which is Eq. (7) of the main text. 

\subsubsection{Phonon-mediated cooling}

In the phonon case, the boson response function is proportional to the usual phonon propagator: 
\begin{equation}
g_{\rm b}^R(\q,\omega) = \frac{2 \omega_{\alpha}^2}{\omega_{\alpha}^2-\omega^2}. 
\end{equation}
The non-interacting electronic response function now involves the electron-phonon matrix elements: 
\begin{equation}
g_{\rm e}^R (q,\omega) = -g_s\int_{\rm BZ} \frac{\d \k}{(2 \pi)^2} \sum_{\nu,\nu'}\frac{ |g_{\alpha,\k,\k+\q}^{\nu \nu'}|^2 }{\hbar \omega_{\alpha}}\frac{n_{\rm F} (E_{\k}^{\nu},T_{\rm e}) - n_{\rm F} (E_{\k+\q}^{\nu'},T_{\rm e})}{E_{\k}^{\nu} - E_{\k+\q}^{\nu'} + \omega + i \delta}. 
\end{equation}
We now show that we recover the results of ref.~\cite{Pogna2021} for the electron-phonon cooling rate obtained in a Boltzamann equation framework, if we neglect electron-electron interactions and treat electron-phonon interactions to first order. 
Under these assumptions, Eq.~\eqref{result} reduces to 
\begin{equation}
\mathcal{Q} = \frac{1}{ 2\pi^3} \int \d \q  \int_0^{+\infty} \d \omega \, \hbar \omega [n_{\rm B}(\omega,T_{\rm e}) - n_{\rm B} (\omega,T_{\rm b})] \Im{g_{\rm e}^R(\q,\omega)}\Im{g_{\rm b}^R(\q,\omega)}. 
\label{result2}
\end{equation}
We notice that 
\begin{equation}
\Im{g_{\rm b}^R(q,\omega)} = \pi \omega_{\alpha}^2 [\delta(\omega-\omega_{\alpha}) - \delta(\omega+\omega_{\alpha})]
\end{equation}
and 
\begin{equation}
\Im{g_{\rm e}^R (\q,\omega)} = \pi g_s \int_{\rm BZ} \frac{\d \k}{(2 \pi)^2} \sum_{\nu,\nu'}\frac{ |g_{\alpha,\k,\k+\q}^{\nu \nu'}|^2 }{\hbar \omega_{\alpha}}[n_{\rm F} (E_{\k}^{\nu},T_{\rm e}) - n_{\rm F} (E_{\k+\q}^{\nu'},T_{\rm e})] \delta(E_{\k}^{\nu} - E_{\k+\q}^{\nu'} + \omega). 
\end{equation}
Moreover, upon integration over $\k$ and $\q$ in Eq.~\eqref{result2}, the angle-dependent parts of the electron-phonon matrix elements vanish, and the intervalley phonons become formally identical to the intravalley phonons: we may introduce the valley degeneracy and carry out integrations over a single Dirac cone. Altogether, we obtain 
\begin{equation}
\begin{split}
\mathcal{Q} = 2 \pi g_s g_v\omega_{\alpha} g_{\alpha}^2  &[n_{\rm B}(\omega_{\alpha},T_{\rm e}) - n_{\rm B} (\omega_{\alpha},T_{\rm b})]\dots \\ \dots \sum_{\nu,\nu'}&\int \frac{\d \q \d \k}{(2\pi)^4} [n_{\rm F} (E_{\k}^{\nu},T_{\rm e}) - n_{\rm F} (E_{\q}^{\nu'},T_{\rm e})] \delta(E_{\k}^{\nu} - E_{\q}^{\nu'} + \omega_{\alpha}). 
\end{split}
\end{equation}
If we introduce another delta function, according to 
\begin{equation}
\begin{split}
\mathcal{Q} = 2 \pi g_s g_v\omega_{\alpha} g_{\alpha}^2  &[n_{\rm B}(\omega_{\alpha},T_{\rm e}) - n_{\rm B} (\omega_{\alpha},T_{\rm b})]\dots \\ \dots \sum_{\nu,\nu'} &\int \frac{\d \q \d \k}{(2\pi)^4} \int_{-\infty}^{+\infty} \d \epsilon [n_{\rm F} (\epsilon-\omega_{\alpha},T_{\rm e}) - n_{\rm F} (\epsilon,T_{\rm e})] \delta(E_{\k}^{\nu} - \epsilon + \omega_{\alpha})\delta(\epsilon-E_{\q}^{\nu'}),  
\end{split}
\end{equation}
we recognize the graphene density of states, 
\begin{equation}
\overline{\nu}(\epsilon) = g_s g_v \sum_{\nu} \int \frac{\d \k}{(2\pi)^2} \delta(\epsilon - E_{\k,\nu}) = \frac{2 |\epsilon|}{\pi v_F^2}. 
\end{equation}
Our result then simplifies according to 
\begin{equation}
\mathcal{Q} = \frac{2 \pi \omega_{\alpha} g_{\alpha}^2}{g_s g_v}  [n_{\rm B}(\omega_{\alpha},T_{\rm e}) - n_{\rm B} (\omega_{\alpha},T_{\rm b})] \int_{-\infty}^{+\infty} \d \epsilon [n_{\rm F} (\epsilon-\omega_{\alpha},T_{\rm e}) - n_{\rm F} (\epsilon,T_{\rm e})] \overline{\nu}(\epsilon) \overline{\nu}(\epsilon - \omega_{\alpha}),
\end{equation}
which is Eq. (18) in the supplementary information of ref.~\cite{Pogna2021}.